\documentclass[12pt, a4paper]{article}

 \usepackage[bottom]{footmisc}

\usepackage{graphicx}
\usepackage{amsthm}   
\usepackage{amsmath} 
\usepackage{amssymb}  
\usepackage{mathrsfs} 
\usepackage{stmaryrd} 
\usepackage{txfonts} 

\usepackage{hyperref}  
\usepackage[capitalize]{cleveref}
\usepackage{setspace}
\usepackage{enumerate}
\usepackage{appendix} 
\usepackage[all]{xy}

\usepackage{hyperref}

\newcommand{\mathsym}[1]{{}}
\newcommand{\unicode}[1]{{}}

\usepackage{color}  
\RequirePackage[dvipsnames,usenames]{xcolor}

\usepackage[obeyFinal]{todonotes}
\usepackage{ifthen,xspace}

\newcommand{\be}{\begin{equation}}
\newcommand{\bel}[1]{\begin{equation}\label{#1}}
\newcommand{\qe}{\end{equation}}
\newcommand{\ee}{\end{equation}}
\newcommand{\eeq}{\end{equation}}
\newcommand{\ba}{\begin{eqnarray}}
\newcommand{\ea}{\end{eqnarray}}

\def\bal#1\eal{\begin{align}#1\end{align}}
\def\bann#1\eann{\begin{align*}#1\end{align*}}

\date{}

\begin{document}

\title{What can we know about that which we cannot even imagine?}

\author{David H. Wolpert \\
 Santa Fe Institute, 1399 Hyde Park Road, Santa Fe, NM, 87501\\
\texttt{http://davidwolpert.weebly.com}
 }

\maketitle

\begin{abstract}
In this essay I will consider a sequence of questions. 
The first questions concern the biological function of intelligence in general, and
cognitive prostheses of human intelligence in particular. These will lead into questions concerning 
human language, perhaps the most important cognitive prosthesis humanity has ever developed. While it is traditional to rhapsodize about the cognitive power encapsulated in human language, I will emphasize how horribly limited human language
is – and therefore how limited our cognitive abilities are, despite their being augmented with language. 
This will lead to questions of whether human mathematics, being ultimately formulated in terms of human
language, is also deeply limited.  I will then combine these questions 
to pose a partial, sort-of, sideways answer to the guiding concern of this essay:
what we can ever discern about that we cannot even conceive?
\end{abstract}

To simplify the pedagogy, I structure this essay as a sequence of questions, each question leading to the next, 
like successive chapters in a monograph. The first questions concern the nature of
intelligence in biological systems, viewed from a long time-scale, evolutionary perspective.
These questions lead into arguments that the
cognitive capabilities of we humans are extraordinarily limited. Specifically, I argue that those very aspects of our mathematics and science
that are vaunted as providing power for investigating the nature of reality
are rather the opposite: they establish that our mathematics and science is actually extraordinarily constrained, a straight-jacket on our
ability to interrogate reality. These arguments then lead to questions concerning
whether our cognitive abilities might forever be limited this way. I then consider 
the implications such limitations would impose, on what we might ever be able to glean concerning
what it is that we will never even be able to imagine. In particular, I consider how future development of AI
might impact this issue.

The first of these questions are similar to questions considered previously, e.g.,~\cite{mcginn1994problem}. However,
the ending questions are novel, at least in their emphasis. 

All of these questions concern the possibility of cognitive constructs whose capability is
beyond that of human minds. Unfortunately, I am a mere
 human (as are most readers of this essay, I presume). So my language in formulating and discussing 
these questions will necessarily be allusive and  somewhat imprecise. Nonetheless, I hope the essence of what I have to say 
is clear. To that end, I finish this essay with suggestions for how
to empirically answer the questions I raise, thereby (I hope!) rescuing this essay from the swamps of 
self-important bloviation to be found in the libraries of humankind under the moniker of ``philosophy''.


To begin this journey of questions, note that high levels of abstract intelligence, high ability to reason about issues other than how
to manage our physical bodies, seems to have been exceedingly rare across all Terran species.\footnote{As with so very many
terms in this essay, I will not engage in the fraught exercise of trying to give a precise definition to the term ``intelligence''.
As mentioned, of necessity, this essay is allusive --- and I trust that the reader gets my drift, even without pages and pages
of tortured text squirreling through subterranean tunnels
trying to gain a grain or two of increased terminological precision.} This is true even if we consider the entire
 history of life on Earth, rather than restrict attention to the set of species currently inhabiting our biosphere.

Indeed, for vast stretches of time the highest level of intelligence among all species
on Earth seems to have increased very slowly, at best. Even if there were not an immediate 
adaptive fitness benefit to a species for increasing its intelligence level, one 
might expect that the random exploration inherent in biological evolution 
would have resulted in a stochastic ``drift" increasing the maximal level of intelligence over all species in the biosphere, much like the drift
d process that played a part in the increase in the complexity of the biosphere as a whole~\cite{krakauer2011darwinian}.

One way to investigate this peculiar reluctance of intelligence to increase
is to adopt a functionalist, teleological perspective on evolution, i.e., to view evolution
as a process that creates increasingly ``optimized'' designs of
 biological organisms. It may simply be that it may takes $\sim$ 
a billion years to design a brain with our levels of intelligence, if the design is done
 using the kinds of ``optimization algorithm" embodied in natural selection. 
Human brains are physically complicated devices. Presumably much of
 that complexity is necessary in any system that has the cognitive capabilities of our brains. 
And presumably, such complexity requires a while
 to evolve under the optimization algorithm embodied in (this very simplified formulation of) natural selection.

While it’s hard to address this possible explanation quantitatively, 
it seems implausible. No other significant capability of biological organisms --- and there are some capabilities that are implemented with very complicated designs --- took a billion years of tinkering to evolve. Other biological capabilities, ranging from ability to navigate through an environment based on patterns of photons reflected from that environment, to the ability to defend against a vast range of microbial assaults, to the ability to disentangle multiple simultaneous sets of coupled waves of atmospheric disturbances 
impacting our ears, and thereby solve ``the cocktail party problem"; all of these abilities involve very complicated mechanisms. And yet all 
have been extremely common across Terran species, both at present and in the past. So this explanation seems to
have major problems.

All other possible explanations for the rarity of high levels of abstract intelligence that I know of have their own
problems  ---  
except for the simplest explanation, the one suggested by Occam’s razor. This explanation follows from the hypothesis 
that the fitness costs associated with high levels of abstract intelligence are large compared to 
the typical associated fitness benefits. It may simply be that evolutionarily speaking, it is stupid to be smart.

Indeed, it seems that at least the sensory-motor information processing in our brains involve all kinds of algorithmic shenanigans so that we can actually use our brains as little as possible, so that we can think as little as possible. Phrased in terms of a sloppy and hackneyed metaphor, it seems that the software running on our brains is designed so that it involves executing
as few lines of code as possible. (See for example the vast literature on the brain's use of
predictive coding~\cite{aitchison2017or,spratling2017review} and sparse coding~\cite{olshausen2004sparse,spanne2015questioning}, 
as well the literature on the low-level processing in the visual system~\cite{rudd2017lightness,bradley2008velocity,van1992information},
in addition to the literature on the coding of more abstract concepts in the brain~\cite{andersen1985encoding,esteves2021spatial}.) 

It is often argued that the underlying reason for this aversion to thinking is to reduce the associated fitness costs~\cite{balasubramanian2015heterogeneity,sterling2015principles}. 
Indeed, such costs to thinking are not difficult to find. In particular, it turns out that brains are extraordinarily expensive metabolically on a per-unit-mass basis, far more than almost all other organs (the heart and liver being the sole exceptions --- see~\cite{bullmore2012economy,sterling2015principles,levy2021communication,balasubramanian2021brain}). 
Consistent with this, it is not just that the software comprising our minds that seems tailored to reduce metabolic costs; the hardware supporting that software   ---  the physical architecture of our brains  ---   also seems tailored to reduce metabolic costs.

We do not have a good understanding of exactly \textit{how} our hardware is used to provide the ability of humans to engage in activities requiring high levels of abstract intelligence. We do not understand how ``brain makes mind''~\cite{bassett2011understanding}.
Indeed, the brain seems to incur major energetic costs even when it is in the resting state, not currently 
engaged in any ``task-focused" activities, but simply providing the ability to engage in such activities~\cite{raichle2015brain}.
In addition, there are many subtleties in how the ``size'' of the brain (which can be quantified many ways) is related
to the associated computational abilities~\cite{herculano2016human}.
Nonetheless, it seems clear that less metabolic cost $\Rightarrow$ less brain mass (smaller neurons and / or number of
neurons, perhaps proportional to other attributes) $\Rightarrow$  less abstract intelligence. 

In light of this advantage of low intelligence, one would expect humans to have the minimal possible computational abilities sufficient for surviving in the particular ecological niche in which we were formed (namely, the niche of social omnivores that inhabit African mixed forest / savanna). One would expect our computational abilities to be extremely limited. 
%
Indeed, make the simple assumption that there are \textit{some} cognitive abilities with the following two
attributes: i) they are not directly helpful for
surviving in our ecological niche (and are also not the inevitable side-effects of some such abilities that \textit{are} helpful);
ii) they are crucial to formulating a science and mathematics (hereafter abbreviated as SAM) that {fully} (!!!) encompasses
the nature of physical reality. Then due to the adaptive fitness costs of all cognitive abilities,
we would have to predict that we do not have these kinds of abilities. In other words,
we would have to conclude that the actual SAM that humanity has created
(or would ever be able to create) cannot fully encompass the nature of physical reality, simply because
natural selection has ensured that we do not have the cognitive abilities necessary to do so.

Biology also provides some purely statistical reasons, not involving arguments based on Darwinian natural selection, which 
suggest that our cognitive abilities are deficient. As a first example, assign a bit to every {other} species on Earth
that we can, a bit that it equals $1$ if we are sure that 
that species has all the cognitive abilities 
necessary to fully apprehend the nature of physical reality. 
There is one species we cannot assign such a bit to: ourselves, since we do not know whether we have deficiencies
in this kind of reasoning. But we can assign
it to all the other billions of species that have ever existed on Earth: 
yes, that species is deficient, cognitively, for the simple reason that there are features of our SAM that no
member of that species is cognitively capable of understanding. We can treat all of these bits as a dataset.
If we applied an admittedly crude Bayesian argument to this particular dataset, along the lines of Laplace's law of succession, we would conclude that it is very likely that the one species for which we cannot assign a bit value  ---  ourselves --- is overwhelmingly likely to 
have such deficiencies as well.

As a second example of such an argument combining evolutionary history and simple statistics, 
fix some way of formalizing a ``type" of cognitive capability. The ability to sense the external world would be one such ``type" of cognitive 
capability; the ability to remember past events would be another; the ability to plan a future sequence of actions would be another, etc. 
The suite of cognitive attributes that we collectively possess that enable our SAM --- our ``intelligence'', if you will--- is yet
another type of cognitive ability. Given any 
such formalization, we could consider the union of all kinds of cognitive capabilities possessed by all organisms on Earth at any given time. We can 
then conduct a time-series analysis of how the size of that union of all kinds of cognitive capabilities of all species on Earth has changed over 
evolutionary history.

As described above, the highest level of intelligence {of any single species} yet reached on Earth is small
(with the single exception of us, perhaps). Yet there are very many other types
of cognitive ability. Arguably, given this breadth of types of cognitive ability, any single species (including
us) has only a small fraction of all the types. Nonetheless,
no matter what precise time-series analysis technique we use, and no matter how we formalize ``type of cognitive capability," 
it seems we will conclude that there is currently a strictly increasing trend in the size of the \textit{union over all species} of their cognitive
capabilities. After all, in no evolutionary period (outside of the several mass extinctions) has the set of all types of cognitive capabilities 
held by any entity in the terrestrial biosphere \textit{shrunk}. The biosphere
as a whole has never lost the ability to engage in certain kinds of cognitive capability. 

So on the one hand, there seems to have been growth
over evolutionary time in the maximal {degree} of the kind of intelligence needed to create SAM of any single terrestrial species.
But there has also been growth in the set of \textit{kinds} of cognitive capability possessed by all species collectively. If we
simply extrapolate either of these trends into the future, we’re 
forced to conclude that there are likely to be levels of intelligence and kinds of cognitive capabilities that some future
organisms will have but that no currently living Terran species has --- including us. 

We can even speculate about just how these future organisms will develop these new kinds of intelligence.
As was first emphasized by Szathmary and Maynard-Smith~\cite{szathmary_major_1995},
the major evolutionary transitions (METs) in terrestrial biology have often involved
jumps to a more collective level of organization, where what used to be the
autonomous, ``individual intelligences'' are subsumed as subsystems
of new, larger ``unit of intelligence''. These METs has been accompanied by what appear
to be jumps in the \textit{kinds} of intellection the new unit is capable of. For 
example, the MET from an individual cell like a paramecium to
the first multicellular organism was accompanied by a
jump in the kinds of computation that the units of intelligence could do.
Similarly, it is a canard that the collective intelligence of a human society exceeds that
of the individual humans within the society. Of course, formalizing a generally applicable
way to specify what the ``units of intelligence'' are in any given complex system like
the terrestrial biosphere is a very fraught issue. But nonetheless, there does seem to
be this loose kind of progression. The important point being that it will presumably 
continue into the future, resulting systems that may be loosely related to the
fantasies of boiler-plate SF, like ``the borg'', and that are far more intellectually capable than we are.

Given all these reasons for supposing we have deficiencies in our cognitive abilities, 
it should not be surprising that the list of our computational limitations is legion. To give one striking example, evidently it was not even 
sufficiently beneficial in terms of adaptive fitness for us to have working memory capable of holding more than seven enumerated objects at once.
Needless to say, there are many artificial computational systems that are capable of this ``feat''.
Indeed, modern digital computers, which are exceedingly simple systems in comparison to the complexity of the human brain, vastly outperform 
us computationally in myriad ways, going far beyond the number of enumerated objects in 
their memories~\cite{brown2020language,devlin2018bert,silver2017mastering}. Moreover, the small set of those
cognitive tasks that we can still perform better than our primitive digital computers is substantially shrinking from year to year --- witness
the recent rise of foundational AI models and the ever-expanding list of new abilities afforded
by deep RL algorithms like DeepMind's Alpha-* algorithms.\footnote{It is important to appreciate that the
rate of this shrinkage is significant, notwithstanding the fact that there are, for now, some important cognitive tasks that 
it has been argued elude AI programs~\cite{mitchell2021ai}, and
notwithstanding the research frustrations endured by some the very earliest of AI researchers, sometimes summarized as ``Moravec’s
paradox''~\cite{brooks1991intelligence,minsky1988society,moravec1988mind}.}

Despite all our preening in front of our collective mirror about how smart we are, it seems that we must in fact be very dumb organisms, by any objective criteria, in terms of abstract intelligence. Thanks to the exigencies of natural selection, we must have highly restricted cognitive abilities --- the barest minimum needed to scrape through a few million years of hunting and gathering until we got lucky and stumbled into the Neolithic revolution. It may simply be that the presumption stated at the beginning of this essay, that we ``have high level of abstract intelligence'' is,
objectively speaking, wrong. It may simply reflect camouflaged biases we have on the topic of ``intelligence'',
esteeming the type that we happen to have above all others. 

However, before getting too comfortable in this conclusion reached by considering the time series of intelligence on Earth, there is a peculiarity of 
that time series that begs investigation. The maximal intelligence on Earth was gradually, smoothly increasing until around
50,000 years ago. It seems that at that time there was a substantial 
jump, when modern \textit{homo sapiens} started on the trajectory that would ultimately
produce all of modern science, art, and philosophy. Is this the case? Are we in fact the fresh-born products
of a substantial jump in the cognitive ability of organisms on Earth?

If we provisionally suppose that there was indeed a major jump in our cognitive abilities very recently, we
arrive at the first major question of this essay:
\begin{enumerate}
\item[1.] 
 Why is there a major chasm, with the minimal cognitive capabilities necessary for survival by pre-Holocene hominids on one side, and on the
 other side, all those cognitive capabilities that Kurt Gödel, Albert Einstein, and Ludwig van Beethoven called upon when conjuring their wonders?
\end{enumerate}

Why is there the major chasm between the computational skill set needed to avoid the proverbial lion lurking in the bush, and the computational skill set that Lady Murasaki, Vyasa, and Al Khwarazmi employed when they wished upon their separate stars?
As the canard goes, there is no evident fitness benefit for a savanna-forged hairless ape to be able to extract from the deepest layers of physical reality cognitive palaces like the Standard Model of particle physics, Chaitin’s incompleteness theorem, or the Zen parable, ``Ten Verses on Oxherding." And in fact, by the arguments above, there are likely major fitness \textit{costs} to our having such abilities.
So why do we have them?

We can gain some perspective on this question by reformulating it more concretely, in the context
of the other species in the terrestrial biosphere:

\begin{enumerate}
\item[2.] Restricting attention to what are, in some sense, the most universal of humanity’s achievements, the most graphic demonstrations of our cognitive abilities: 

\textit{Why were we able to construct present-day science and mathematics, but no other species ever did?} Why have we -- uniquely -- 
been able to decipher some features of the Cosmic Baker's cooking techniques, by scrutinizing the breadcrumbs that 
they scattered about the universe? Why do we
 have that cognitive ability despite its fitness costs? Was it some subtle requirement of the ecological niche in which we were formed --- a niche
 that at first glance appears rather pedestrian, and certainly does not overtly select for
the ability to construct something like quantum chronodynamics? Or is our 
cognitive ability a spandrel, to use Gould and Lewontin’s famous phrase --- an evolutionary byproduct of some other trait? 
In other words, is it all just a cosmic fluke?
\end{enumerate}

This is hardly a new question, of course. But considered carefully,
it gives rise to a sequence of many other, less commonly considered questions.
To begin:

\begin{enumerate}
\item [3.] Are we really sure that no other species ever constructed some equivalent of present-day SAM? 
Are we really sure that no other apes --- or cetaceans or cephalopods --- have achieved some equivalent of our SAM, but an equivalent that we are too limited to 
perceive?
\end{enumerate}

I would say that this question at least we can answer,
and the answer is `yes'. It may well be that, if and when we come to partially decipher the communication
systems of some non-human species~\cite{andreas2021cetacean}, 
we will find messages in their communications that in some sense correspond to our SAM. For argument’s sake, 
let’s suppose that this \textit{will} be the case --- that we will find such science(ish) and math(ish) messages in the communications among
members of some other species. 

Even with that assumption, we would still have to conclude that the SAM of all other terrestrial species
is inferior to ours in some extremely important respects.
After all, incontrovertibly, the Earth’s physical environment reflects the effects of \textit{our} understanding of SAM --- it reflects our activities --- far
more than it reflects the activities of any other single species~\cite{bar2018biomass}.
There is no reason to believe that other species would not have also completely transformed the biosphere to serve their ends, just as we have, if only they had the SAM that is necessary to do so. The implication, given the absence of such a transformation by some other species, is that no such species has developed as sophisticated a SAM as we have.

There is nothing subtle about this evidence of our singular abilities in SAM. Nor is reasoning based on this evidence
particularly biased in our favor. 
After all, the evidence of our abilities in SAM, given by our transformation of the biosphere, is obvious to members of other species (many of
whom we’ve actually wiped out), whereas there is no evidence of \textit{their} abilities in SAM that is similarly obvious to us.

This answer to question (3) suggests that we should modify question (2):

\begin{enumerate}
\item [4.] If the evidence of the uniqueness of our SAM is the modifications that we, uniquely, have wreaked upon the terrestrial biosphere, should the question really be
why we are the only species who had the cognitive abilities to construct our SAM \textit{and} were able to build upon that understanding, to so massively re-engineer our
environment? To give a simple example, might some cetaceans even exceed our SAM, but just do not have the physical bodies that would allow them to exploit that understanding to re-engineer the biosphere in any way? Should the focus of the inquiry not be whether we are the only ones who had the cognitive abilities to construct our current SAM, but rather should the focus be expanded, to whether we are the only ones who had both those abilities \textit{and the ancillary physical abilities} 
(e.g., opposable thumbs) that allowed us to produce physical evidence of our SAM?
\end{enumerate}

Note that the combination of our SAM and our physical ability to exploit that SAM has not 
only provided us with the ability to twist the biosphere to satisfy our desires, but also 
with cognitive prostheses~\cite{ford1997other} and extended minds~\cite{clark2012extended,colombo2019andy,clark2007re}. 
Furthermore, the capabilities of those extended minds have been greatly magnified over time
--- by the cumulative collective process of culture and technological
development~\cite{derex2019causal,erwin2004insights,henrich2016secret}. In turn, 
these extended minds have accelerated the development of culture and technology.
There is a feedback loop between the collective capability of our cognitive prostheses and our technology.

This feedback loop has allowed us to ``build out" the set of our original cognitive capabilities, to expand its breadth in an exponentially growing
 dynamics that is not solely driven by genotypic evolution. This distinguishes the current expansion in our capabilities
from their expansion before the
Pleistocene, and from all expansions in the cognitive abilities for all other Terran species, which were driven by genotypic evolution.\footnote{I'm
here ignoring the cultural evolution that seems to take place in some primates and Cetaceans, as being inconsequential
to the issues considered in this essay.} 
Concretely, this combination of our high current cognitive abilities and the physical ability to exploit
them may not only have allowed us to expand our SAM beyond whatever other Terran species have already achieved,
but even beyond what
other Terran ever \textit{can} achieve. Indeed, it may have been this feedback loop, between the abilities afforded by
our cognitive prostheses and our technology, which allowed the gap to form between
our cognitive capabilities and the minimum such abilities necessary for survival on the Veldt. In other words, this loop may
provide the answer to question (1).

However, a perhaps crucial feature of the feedback loop is that while it has inflated our \textit{original} cognitive capabilities, it is not clear that it has provided us any 
\textit{wholly new} cognitive capabilities. In fact, it might not ever be able to. The future phenotypic possibilities of any species evolving according to genotypic evolution are constrained by the frozen accidents in its evolutionary past~\cite{aguirre2018networked,cracraft2004assembling,manrubia2021genotypes}, 
which limit its possible future forms. This may not only be true for relatively slow genotypic evolution; it may be just as true if the evolution is very fast and driven by a feedback loop between extended minds and collective culture.
Perhaps our future SAM will be constrained by the set of cognitive capabilities we had when we started to construct our current SAM. 

This suggests a different kind of resolution to question (1). Maybe the gap between our
current SAM, on the one hand, and the kinds of knowledge our minds were designed to uncover by traipsing
across the infamous Veldt, on the other hand, perhaps that gap is not a ``chasm." Perhaps it is more accurate to describe 
the gap as a small divot, with the full range of all that will forever be unknown to us, 
all that our SAM can never encompass, extending far beyond our current position, on this
side of the gap. 

Yes, sure, perhaps our SAM and our ability to exploit our SAM far exceeds that of other 
species on Earth. But perhaps these are tiny differences, almost invisible
on the scale of what would accompany cognitive capabilities that are more extensive than those that we 
will ever have. Maybe the answer to Wigner's famous question, about why our mathematics is
``unreasonably effective'' at capturing the nature of our physical reality~\cite{wigner1990unreasonable}, is simply that 
our mathematics \textit{isn't very effective at all}, that in fact 
our mathematics can only capture a tiny sliver of reality. Perhaps the only reason it \textit{appears} to
be so effective is because our vista is so restricted, to just those very, very few aspects of reality
that we can conceive of.\footnote{See~\cite{hamming1980unreasonable} for similar sentiments
concerning Wigner's question.}

Therefore, perhaps the interesting question is not why our augmented minds seem to
have abilities greater than those necessary for the survival of our ancestors, but rather whether
our augmented minds have the minimal abilities necessary for grasping reality:

\begin{enumerate}
\item [5.]	Ancillary abilities or no, are we unavoidably limited to enlarging and enriching the SAM that was produced by our species with the few cognitive abilities we were
born with? Is it impossible for us to concoct wholly new types of cognitive abilities --- computational powers that are wholly novel \textit{in kind} --- which in turn could
provide us wholly new kinds of SAM, kinds of SAM that would concern aspects of physical reality currently beyond our ken?
\end{enumerate}

Perhaps the earliest versions of this question in modern, Western literature were throwaway phrases scattered among various essays. In these early, verbal baubles it was suggested that the universe may be ``{queerer / stranger / odder} than we can {suppose / imagine / conceive}"~\cite{haldane1928possible,clarke2013profiles}.
Having other fish to fry, the authors of these early texts rarely fleshed out what they meant with these phrases. 
In fact, they often seemed to mean that the universe may be stranger than we can \textit{currently} imagine, due to limitations in current scientific
understanding, rather than being stranger than we can \textit{ever} 
imagine, with any future science, due to inherent limitations of what we can ever do
with (future efflorescences of) our minds.\footnote{For example, in \cite{clarke2013profiles}, Clarke
explicitly considers the first type of strangeness, not the second,
when he introduces the phrase, while in \cite{haldane1928possible}, Haldane
explicitly predicts that we will ``one day … be able to look at existence from the point of view of non-human minds", which seems to rule out the second type of strangeness.}

Various forms of question (5) were then raised later with a bit more care and detail~\cite{barrow1999impossibility,chomsky1997language,fodor1983modularity,koppl2011hayek,pinker2003mind}. 
However, most of those later forms of the question have concerned the dubious ``hard problem of consciousness"~\cite{harris2006reflections} and the closely related ``mind --- body problem"~\cite{mcginn1994problem}. They did not concern the issues under consideration
in this essay.

Fortunately, we can approach question (5) in a more rigorous manner.
To see how, we can start with the recently (re)popularized idea that some portion of our physical
universe, including in particular we denizens of that physical universe, might just be 
part of a simulation produced in a physical computer of some super-sophisticated
race of aliens~\cite{bostrom2003we,hamieh2021simulation,chalmers2022reality+}.
In other words, under this ``simulation hypothesis'', we would just be a computer program running in a physical computer of such
a putative super-sophisticated race of aliens.\footnote{The reader should be warned that the term ``simulation''
actually has a formal definition in the computer science theory of state transition systems~\cite{wikipedia_state_transition_system},
which differs somewhat from what ``simulation'' means in the context of the simulation hypothesis.}

This idea can be extended in an obvious manner, simply by noting that the
aliens that simulate our universe might themselves be a simulation
in the computer of some even more sophisticated species, and so on and so on,
in a sequence of ever-more sophisticated aliens. Going the other direction,  in the
not too distant future we might produce our own simulation of a universe, complete with entities that have ``cognitive capabilities''.
Indeed, we might produce such a simulation whose cognitive entities can produce their own
simulated universe in turn, etc. So there might be a sequence of species', each
one with a computer running a simulation that produces the one just below
it in the sequence, with us somewhere in the middle of that sequence.

The question considered in this literature is whether we \textit{are} such a simulation.
It is answered rather trivially if we adopt the view of ontic structural realism, especially if 
that view is formalized in terms of Tegmark’s level IV multiverse~\cite{tegmark1998theory,tegmark2008mathematical}: 
yes, in some universes we are a simulation, and no, in some other universes we are not. 
(See also~\cite{carroll_wilczek_simulation_universe_2021,piccinini2011physical}.)


To make things more interesting though, let's consider what the consequences would be if we are indeed a simulation. 
There are many rich mathematical issues this would raise. 
Consider all possible sequences of \{species running a
computer that simulates a species running a computer that simulates ...\}.
Formally, each such sequence is a path in a directed \textbf{simulation} graph.
What is the form of the full set of such simulation graphs?
Is it a partially ordered set, perhaps a lattice, or just a linear order, or other some other kind of directed graph,
possibly involving cycles? Are there restrictions on the fan-in and / or fan-out of the nodes in that ``simulation graph''?

However, perhaps the most important of these questions for this essay is the following: 
\begin{enumerate}
\item [6.] Are there unavoidable restrictions on the computational power of a species being simulated in a computer, 
in comparison to the computational power of the species running the simulation? In particular, might it necessarily
be the case --- due to the very meaning ``simulation'' --- that there
are questions that can be answered by the simulating species, but that are impossible to answer for the species 
being simulated?
\end{enumerate}

Strangely, even though the simulation hypothesis has always been discussed in terms of a \textit{computer} simulating
the dynamics of our universe, nobody has investigated it before using the tools of computer science theory, or of
its cousins like logic theory.\footnote{Though an 
attempt was made  in~\cite{campbell2017testing} to investigate the simulation hypothesis using concepts from uncertainty quantification, which
in certain respects is related to computer science theory.} 
However, there are many ways to investigate question (6) that are grounded in computer science theory. One approach resulting in theorems
having some strange implications for philosophy
is based on Kleene's second recursion theorem~\cite{wolpert_2024_self-sim}. But there are many other
approaches as well.

One of those other approaches begins by picking any formal system that is powerful enough
to be subject to G{\"o}del's two incompleteness theorems. The first of those two theorems
establish that certain mathematical statements statable in that formal system
cannot be proven in that formal system (i.e., certain ``well-formed formulas'', to use the
proper terminology, cannot be proven in that formal system).
Such a statement in the formal system is ``true'', loosely speaking --- but cannot be proven 
using that formal system.

Now loosely speaking, all one needs to do to nullify this conclusion of the first theorem
is to introduce an extra axiom into the formal system that states whether the 
mathematical statement that is causing trouble is true or
not. Bang --- the apparently paradoxical statement giving the incompleteness
theorem has been resolved. 

Unfortunately though, the combination of the original formal system
with that new axiom is, strictly speaking, a more powerful formal system than the original one. Moreover,
\textit{that} new, more powerful formal system is itself subject to the first incompleteness theorem of G{\"o}del, 
just referring to a different mathematical statement (i.e., a different ``well-formed formula'', to use
proper terminology) to the first use of the theorem. And of course, this
conclusion concerning that new statement in that new formal system could itself be resolved by introducing 
yet another axiom, resulting in a yet more powerful formal system~\cite{barrow2011godel}. 
This sequence of increasingly powerful systems has no upper limit --- each new formal
system is strictly more powerful than the ones below it whose incompleteness theorems
it obviates. 

In fact, the \textit{second} incompleteness theorem of G{\"o}del proves that there
is no way to prove that a sufficiently powerful formal system is consistent and can prove
all true mathematical statements one could state in terms of that formal system.
No matter how we augment a proof in a formal system, no matter how we combine it with
other proofs in that formal system, there will be proofs in the formal systems higher in the hierarchy
that it can never replicate.\footnote{Interestingly, while there is no upper limit to this sequence of formal systems / simulations,
there \textit{is} a lower limit; it is the formal system that arises in G{\"o}del's incompleteness
theorem, whose axioms include those of Peano's arithmetic. (In weaker formal systems it is not even be possible to perform Peano
arithmetic operations.) 

In light of this, suppose we
identify each member of a sequence of increasingly powerful formal systems, able to prove theorems concerning
the formal systems lower down in the sequence than it is, with one of the ``computer simulations 
produced by sophisticated aliens'' in the nested-dolls sequence of simulations considered  in~\cite{bostrom2003we}.
Under this interpretation of the term ``simulate'', there is an infinite set of ``simulations'' that are strictly
more computationally powerful than the one giving our physical universe --- but none that we would consider
computationally interesting that are weaker.} In this sense, \textit{no matter how our physical brains are organized},
and \textit{no matter what physical systems we ever construct to augment our brains}, we
will be cognitively weaker than the aliens residing in those other, more powerful formal systems,
peering down at us (assuming of course that
the cognitive abilities of those aliens includes (dis)proving all mathematical statements in the formal
system that defines us, including in particular all statements which which we cannot (dis)prove).

Yet another formalization of question (6) involves various 
``reduction hierarchies'' (arithmetic and analytic hierarchies, and  
their resource-limited forms, the
polynomial and exponential hierarchies~\cite{cooper2017computability,hajek1979arithmetical}). 
The computer science form of these hierarchies involve
decision problems which we wish a Turing machine to solve.\footnote{More formally, they involve bit strings codifying decision problems that must be assigned either the value `true' or the value `false'
by running an appropriately programmed Turing machine that takes that bit string as input.
See~\cite{arora2009computational}.} Very loosely speaking, define
some set of (appropriately formalized) decision problems, $\alpha$. Next choose a set $A \subset \alpha$
that can be decided by a Turing machine subject to a certain bound on their time complexity (e.g., polynomial complexity
P or non-deterministic complexity NP). Informally, this means that none of the decision problems
in $A$ become much harder to solve as the precise instance of that problem grows larger and
larger, as far as that Turning machine is concerned. In contrast, instances of decision problems lying outside
of $A$ get to be extremely difficult to solve as the instances grow larger. No matter how we program
that Turing machine, no matter how we ``augment'' one program running on it with another one,
it will always find problems outside of $A$ to be far more difficult to solve than those inside of $A$.
%

Next, define a larger set of problems, $B \subset \alpha$, which contains all the problems in $A$, and which can be
decided with the same time complexity as those in $A$ (i.e., that are ``just as difficult to decide'' as those
in $A$), by now allowing that same Turing machine to call an ``oracle'' 
that can instantly decide whether a given problem is in $\alpha$ or not, and do so an arbitrary
number of times. This new, souped-up Turing machine, with access to the oracle, is strictly
stronger than the original Turing machine, in the simple sense that there is a larger set
of decision problems, containing the original set, that are all easy for 
the new Turing machine. (In contrast, problems in $B$ that are outside of $A$ are strictly
harder to decide for the original Turing machine, which did not have access to the oracle.)

We can iterate this ``oracle-izing'' procedure, by defining a new oracle that can
instantly decide whether a given problem is in $B$. 
By allowing our original Turing machine
to call this new oracle an arbitrary number of times, we allow it to decide a
set of problems, $C$, which is even larger than $B$, again with the same time
complexity as the original Turing machine decided the original set of problems $A$.
In this way we define an entire infinite ``hierarchy'' of increasingly powerful Turing machines,
able to decide an increasingly large set of decision problems without expending extra effort.
Crucially, no matter how we write the programs of the Turing machine at one level $k$ in the hierarchy,
no matter how we ``augment'' it with other programs run on the same machine, there will
always be problems that are easy  to solve for the Turing machines at levels in the hierarchy higher than $k$,
but are effectively impossible for any Turing machine at level $k$.

There are other more powerful ways to exploit oracles however. For example, the Turing degrees of
computability theory involve oracles that provide answers to decision problems that are formally undecidable
to a computational machines with lower Turing degree~\cite{soare2016turing,shore2016turing}. For example, the computational machines
with Turing degree $0$ are just conventional Turing machines, while the machines with Turing degree $1$
are the conventional Turing machines augmented with an oracle that can decide the Halting problem for
all conventional Turing machines. Again, we get an hierarchy of increasingly powerful computational machines,
in this case able to decide an increasingly large set of decision problems that cannot be decided by the machines
lower in the hierarchy even if those lower machines expend an arbitrarily large amount of effort.

These arguments involving hierarchies of increasingly powerful computational machines,
able to decide ever larger sets of decision problems with the same amount of effort,
is analogous to the argument above concerning a sequence
of increasingly powerful formal systems, each of which
resolves the impossibility theorems of the formal systems below it in the sequence
but is weaker than the formal systems above it. 
The implication are analogous as well: no matter
how we augment our minds --- no matter where we lie in the infinite sequence of ever more powerful
oracles --- we will be more limited in our cognitive abilities
than the minds higher in the sequence.\footnote{More precisely, no matter how we program a Turing machine
to exploit whatever oracles our minds have access to, there will be other Turing machines, able to access other
oracles, that once they are properly programmed
can solve a larger set of decision problems than we can for the same computational effort.}
In summary, if we make these other definitions of the words `simulate' and `computer', 
the resultant answers to question (6) strongly suggest that augmenting our brains 
can never allow us to fully grasp / cognize / perceive our physical reality.

All of these arguments concerning question (6) are some examples of the \textit{content} of our mathematics that suggest that we humans
are too limited in our cognitive abilities to cognitively engage with reality in its full extent.
Some of these arguments, e.g., the ones invoking G{\"o}del's incompleteness theorems, 
have been grappled with in the literature for close to a century. 

However, there are other aspects of our mathematics that
instead involve its \textit{form}, not its content. These arguments 
also strongly suggest that we are highly constrained in our cognitive capabilities. Unfortunately, these aspects of mathematics have
received far (if any) consideration in past literature. I turn attention to these next.

One of the most powerful fields of mathematics to develop in the past half century is category theory~\cite{awodey2010category,riehl2017category}. Category theory provides an extraordinarily 
concise way of unifying almost all fields of (human) mathematics, ranging from set theory to algebra to
topology to analysis. By
stripping out the extraneous details, it makes glaringly obvious that all
our fields of mathematics are just variants of the same underlying structures, known as ``categories''. What makes
category theory so {beautiful} is that the definition of
a category is almost absurdly simple: a category consists of abstract ``objects'', together
with ordered pairs of objects, which are called ``morphisms''. There are a few simple requirements imposed on how
the morphisms in a category must be related to
one another. \textit{And that is it.} All of our mathematics reduces to embellishments of these simple structures.

One response to category theory ---  
part of the reason it strikes so many people as beautiful --- is to be stunned by it. Category theory shows us
that the many fields of human mathematics, which we are used to viewing
as many different types of flower filling a great garden, are in fact many instances of
the same underlying, singular plant, planted all over our yard,
simply viewed from many different angles. This response might lead one to suppose that the simplicity of category theory
reflects some kind of simplicity of mathematical reality. 

However, being a bit more objective about things,  it might just as well be that the
simplicity of the mathematical garden we have tilled reflects a limit of \textit{us}, the gardeners,
not of the number of species of mathematical plant there are. It may be that 
the reason that our fields of mathematics are all embellishments of a simple underlying structure
is that they were created by simple-minded beings: us. It might be that
there is a vast expanse of mathematical reality which can\textit{not} be expressed in such an extremely simple form, but is intrinsically  
more complex --- and is forever beyond us.

This leads to the following question:

\begin{enumerate}
\item [7.]	Is the very \textit{form} of the SAM that we humans have created severely constrained somehow? 
\end{enumerate}

One does not have to learn category theory to address this question.
If one steps back from our SAM, viewing it from afar, one notices that there are striking limitations in how our SAM is formulated, limitations that are common to the SAM produced by almost all human societies. These are limitations in the kinds of patterns with which our SAM is
\textit{represented}. 

As was explicitly recognized at least as far back as Wittgenstein~\cite{biletzki2002ludwig,schulte1980wittgenstein,wittgenstein2013tractatus}, 
all of human mathematics consists of discrete sequences of clearly delineated patterns composed either of marks on a surface or words in speech. All statements in mathematics consist of finite sequences of elements from a finite set; e.g., ``1 + 1 = 2" is a sequence of five elements from a finite set of symbols. All proofs in mathematics --- all theorems based on Zermelo --- Fraenkel Choice (ZFC) set theory, all predicate logics, all category theory --- comprise a finite discrete sequence of such statements.

Moreover, to paraphrase Galileo, all of our physics, and so all of our understanding of the foundations of physical reality, is  
formulated in terms of our mathematics~\cite{hut2006math,tegmark1998theory,tegmark2008mathematical,wolpert2020noisy}. 
%
Much of philosophy has reacted to this observation, that our SAM is just a set of finite sequences of symbols, 
by trying to unpack / formalize the precise way that such finite sequences can ``refer to something outside of 
themselves."~\footnote{To appreciate some of the difficulty of this problem,
note that any sequence of symbols has no more significance in and of itself about any one specific
mathematical or physical object than do the sequences one might find in the entrails of a sacrificed sheep, or in the sequence of cracks in a heated tortoise shell. How to
circumvent this issue is called the “symbol-grounding problem” in philosophy~\cite{fodor1975language,harnad1990symbol}.} 
The community of mathematicians has reacted in a similar way, expanding formal logic to include modern model
theory~\cite{hodges1993model} and meta-mathematics~\cite{kleene1952introduction}.

Arguably though, what is truly stunning about the fact that modern 
SAM is formulated in terms of finite sequences is its exclusivity; \textit{nothing} other than such finite chains of symbols is \textit{ever} found
anywhere in modern mathematical reasoning. So fundamental is this restriction in current 
mathematical reasoning that it has implicitly been enshrined in our current definition of such reasoning, via the Church-Turing (CT)
thesis~\footnote{
My concern here is not whether the CT thesis is true, or what its implications would be
for whether there can be a physical computer in our universe that simulates the evolution of our entire
universe, including that computer itself. 
My concern now is more fundamental, with whether the implicit assumption of thesis, 
that the physical universe can be formulated in terms of a set of sequences of symbols, is valid.}.

So evidently, the answer to question (7) is `yes'.
This answer immediately gives rise to yet more questions.\footnote{Some of these are outside of the scope of this essay, e.g., whether
there might somehow be significant implications of the fact that, introspectively, we seem to
expend a ``small amount of effort'' when we reason using
human language~\cite{chomsky1997language}.} The most disquieting of these is whether
there might be some reason for these limitations on the form of our SAM that has nothing to do with
what we are trying to achieve with our SAM, but rather reflects how that SAM developed. Might our motives in using such a severely restricted
form of SAM be less than pure, so to speak?
\begin{enumerate}
\item [8.]
Is human SAM formulated exclusively in terms of finite sequences of symbols from finite alphabets
due to some contingent aspect of our evolutionary and / or social history?
\end{enumerate}

Well ... as noted above, the highly limited form of all of human mathematics --- sequences of finite strings of 
symbols --- just happens to be \textit{exactly} the structure that we humans use to converse with one another: the structure of human language. 
Indeed, again in large part due to Wittgenstein, it has now become commonplace to \textit{identify} 
mathematics as a special case of human language, casting its structure explicitly in terms of formal grammars,
in the same sense as such grammars arise in human conversation. Note as well that it is a poetic cliche that libraries, and in 
particular mathematics libraries, are places where we converse (!) with past minds. The implication is that the contents of all the mathematics 
textbooks in those libraries is part of a conversation; i.e., an exercise in human language. Indeed, historically, mathematics textbooks and papers 
developed from written correspondence, i.e., from exercises in human language.

So, the form of human mathematics, and of our SAM more generally, just happens to exactly coincide with the form of inter-human communication. So the answer to question (8) is `yes'.

Some writers have pointed this out before, that human language’s design matches that of formal logic and Turing machine theory~\cite{barrow1999impossibility}. 
They have taken this as a wonderful stroke of fortune, that we just so happen have a cognitive prosthesis --- human
language --- that is capable of capturing formal logic. After all --- they presume --- 
this means we are capable of capturing all the laws of the physical universe.

A cynic might respond to this view with heavy irony, ``Gee, 
how lucky can you get? Humans have exactly the cognitive capabilities needed to capture all
aspects of physical reality, and not a drop more!" This cynic might go on to wonder whether an ant, who is only capable of formulating the ``rules
of the universe" in terms of pheromone trails, would conclude that it is a great stroke of fortune that they happen to have the cognitive capability 
of doing precisely that; or whether a phototropic plant would conclude that it is a stroke of fortune that they happen to have the cognitive 
capability to track the sun, since that must mean that they can formulate the rules of the universe. 


Sure, it’s possible that it is just a coincidence, that for some unknown reason the deepest nature of physical reality is expressible in terms of one of our cognitive prostheses. But it certainly seems as plausible that social computation is simply the 
most sophisticated cognitive prosthesis we have ever developed, and
that even exploiting it to the hilt only allows us to capture a sliver of physical reality. Yes, our science and mathematics --- or more precisely, what they seem to be developing into --- may be a complete description of {what we understand physical reality to be}. They might be developing
into a complete description of what is experimentally accessible to us,  even if only
indirectly, both now and in the future~\cite{tegmark1998theory,tegmark2008mathematical,wolpert2020noisy,wolpert2022stochastic}. 

But in 
an exactly parallel manner, a putative ant-level theory of 
reality in terms of pheromone trails and environmental chemical signals could capture all that ants ``understand physical reality to be'',
of all that ants can ``experimentally access''. 
And just as there is a huge expanse of physical reality lying beyond the charmed sliver that ants can conceive of, it may be that there is a 
huge expanse of physical reality beyond our ability to even conceive of.

After all, social computation --- human language --- was
developed for communal sessions of shooting the shit around the campfire after a successful mastodon hunt (plus a few other purposes). There is no reason to believe that
features well-suited for such exercises in nocturnal braggadocio can also be used to glean substantial insights into the shape of the 
hands of the Cosmic Baker, based solely on some crumbs we have discovered, scattered on their kitchen floor.

Noam Chomsky~\cite{chomsky2014minimal}, Daniel Dennett~\cite{dennett2009darwin} and others~\cite{dennett2001mind} 
have marveled at the fact that human language allows recursion, i.e., that it allows arbitrary finite, discrete sequences of symbols from a finite alphabet. They marvel at the fact that humans can create any of a countable number of sentences, viewing this as allowing an amazingly large set of human languages. 
In contrast, I marvel at how restricted human language is, and so how restricted human SAM is. I marvel at the fact that that limitation appears to be universal.
After all, suppose that it were the case that all of physical reality could in fact be formulated in terms of some finite sequence of 
symbol strings from a finite alphabet. Even so, it would still be striking that, as humans converge to that correct formulation, 
we have never considered formulations that are not representable this way, perhaps due to considering an erroneous SAM, or perhaps due to choosing an inefficient representation of SAM. I marvel at the small size of the sandbox we play in while formulating our SAM.

This marveling at the sizes of sandboxes brings to the fore the reason that question (8) is unsettling:
\begin{enumerate}
\item [9.]
Is it a lucky coincidence that all of mathematical and physical reality can be formulated in terms of our current cognitive abilities, including, in particular, the most sophisticated cognitive prosthesis we currently possess: human language? Or is it just that, tautologically, we cannot conceive of any 
aspects of mathematical and physical reality that cannot be formulated in terms of our cognitive capabilities?
\end{enumerate}

To gain perspective on question (9), recall the discussion above of how the maximal level of intelligence on Earth
has grown over evolutionary time, and how one can run a simple time-series fit to extrapolate that maximal level of
intelligence into our future. Well ... can such a time-series analysis reveal more than just a trend line? Does
it tell us anything about the gaps in the \textit{kinds}
of cognitive capabilities which separate us from  our past, suggesting that those gaps which might have analogs in our future?

In contrast to us, no unicellular Eukaryote
like a paramecium can even \textit{conceive} of the concept of a ``question," concerning issues that have no direct impact on its behavior, despite how obvious that concept seems to us. Not only would a paramecium not understand the possible answers we have considered for our questions concerning reality, it would not understand the questions, as has often been noted~\cite{boudry2018science,boundryvideolimits_2}. 
More fundamentally though, no paramecium can even conceive of the idea of posing a question concerning physical reality in the first place. Insofar as the cognitive concept of questions and answers might be a crucial tool to any understanding of physical reality, 
a paramecium is, by construction, 
lacking the tools needed to understand physical reality. It presumably does not even understand what ``understanding reality" means, 
in the sense that we mean the term. Ultimately, this is due to limitations in the very kind of cognitive capabilities of paramecia. 
And as just argued, we, too, almost 
surely have limitations in the very kind of our cognitive capabilities. 

So one way to address question (9) is to note that it is a \textit{question}. It is one of a sequence of such questions that
have been considered in this essay. Given that single-celled organisms could not even conceive of such a thing as a question,
this leads to a more specific (and ironically self-referential) question:

\begin{enumerate}
\item [10.]
Are there cognitive constructs of some sort, as fundamental as the very idea of questions and answers, that are necessary for understanding physical reality, and that are
forever beyond our ability to even imagine due to the limitations of our brains, just as the notion of a question is forever beyond a paramecium?

\end{enumerate}


It may help to clarify this question by emphasizing what it is not.

This question does not concern limitations on what we can know about what it is that we can never \textit{know}.\footnote{I am using the term
 ``know'' 
loosely here, since the distinctions between its various possible formal definitions~\cite{fagin2003reasoning} are not relevant to my arguments.}
Many things can be conceived by us humans even if they can never be known by us. 
The set of what it is that we cannot even \textit{conceive of} is
a (strictly smaller) subset of what it is that we cannot know.
The issue I am concerned with is what we can ever perceive concerning that smaller set, the set of all that we cannot conceive of.

For example, I am not concerned here with the unknowables of how other branches of the many worlds of quantum mechanics turn out~\cite{everett2015relative}.
Nor am I concerned with values of variables that are unknown to us simply because we cannot directly observe them,
e.g., values of variables concerning events outside of
our Hubble sphere, or events within the event horizon of a black hole, or events like how many cells were in a 
dinosaur named ``Bob" who lived in 90 million BC, spending his whole life right where
my house is now. The first one is calculable, in theory at least, even if not observable, and so is arguably knowable. The remaining three are not reflections of limits of our
cognitive abilities --- we can certainly conceive of those variables and their values without any difficulty. Rather, they can never be known 
by us for the simple reason that
our ancillary engineering capabilities are not up to the task, not for any reasons intrinsic to limitations of the science and math our minds can construct. They can be known,
but we cannot find a path to such knowledge.

Also, I am not concerned here with the limitations of what can be known by humans that arise from the logical nature of self-referentiality, nor the limitations of 
Tarski~\cite{smullyan1992godel} or Rice~\cite{hopcroft1979introduction} or Chaitin~\cite{barrow1999impossibility,chaitin2011goedel}. 
Indeed, the simple fact that we can prove that we cannot have this kind of knowledge means that we can conceive of having it.
The impossibility theorems of G{\"o}del and the computational hierarchies discussed shortly after question (6) are not the issues I wish
to highlight at this point --- after all, I am able to describe 
those kinds of unknowability. The concern here is what kinds of unknowable cognitive constructs
might exist that we can never even be aware of, never mind describe (and never mind implement!).

In addition, I am not concerned with what cannot be known \textit{with complete assurance}, for essentially statistical reasons. For example, I am not here concerned with
the fact we can never be 100\%, cross-my-heart-and-hope-to-die sure that a proof we write down has no logical errors, no matter how many times other people and
computer programs might check it. It is conceivable that this limitation of what is knowable would apply no matter what 
kind of intelligence we had --- whether we were
beings far advanced evolutionarily from \textit{Homo sapiens} or were just lowly paramecia.

Similarly, I am not here concerned with theoretical impossibility of inductive inference. I am not concerned with the fact that if one restricts
attention to future experiments, then, as Hume intuited~\cite{hume2003treatise}, there is no assumption-free way to establish the validity of the
scientific method~\cite{wolpert2021important,wolpert2021implications}.
I am not concerned with the fact that none of machine learning can be formally justified in a non-trivial sense without making 
assumptions~\cite{wolp96ab,wolp96b,wolpert2021implications}, 
that black-box optimization algorithms like simulated annealing cannot be justified without making assumptions~\cite{woma97}, 
or that the use of Monte Carlo techniques to estimate expected values of functions cannot be justified without making assumptions~\cite{wolpert2021important}.

Nor am I concerned with the kinds of impossibility results that would apply to any physical universe, regardless of its laws of physics, so long as entities within that physical universe could both encode questions concerning the state of that universe and answer them~\cite{wolpert2008physical,wolpert2017constraints,wolpert2018theories}. 
Like the paradoxes of Turing computability theory and mathematical logic mentioned above, these impossibilities lie within the remit of current (meta)mathematics rather than outside of it.


I am concerned with --- and I wish to draw attention to --- the issue of whether there are cognitive constructs that we cannot conceive of but that are as crucial to understanding physical reality as is the simple construct of a question. 
The paramecium cannot even conceive of the cognitive construct of a ``question" in the first place, never mind formulate or answer a question;
are there, similarly, cognitive constructs that we cannot conceive of, but that are just as necessary to knowing all of physical reality as is the
simple idea of questions and answers? I am emphasizing the possibility of things that are knowable, but not to us, because we are not capable of
conceiving of that kind of knowledge in the first place.

Before proceeding further, this is a good point in this essay to address the conundrum that --- of course ---
this entire essay is itself a sequence of sentences written in English. My arguments
in this essay are themselves a ``finite sequence of finite sequences of symbols, all those symbols from the same finite alphabet’’. 
But the argument I’m making in this essay is that any argument couched exclusively in terms of sequences of symbols may be deficient, unable to capture all (or even most) of reality. (This essay is, loosely speaking, itself an example of self-reference.) In particular, if one accepts the argument of this essay, one concludes that there may be aspects of reality beyond the remit of that argument. Moreover, 
{\textit{a priori}}, those aspects of reality beyond the remit of the argument in this essay may obviate that very argument. 
Cutting to the nub of it: the argument in this essay argues that that argument might be incomplete.
 
Mathematicians are well versed in such logical conundrums. Gödel’s first incompleteness theorem applies to all any logical system at least
powerful enough to describe simple arithmetic with the integers. The first theorem constructs a ``Gödel sentence’’, $s$, in any such logical
system $T$, a sentence which {can most naturally be interpreted} as saying that $s$ cannot be proven in $T$. Colloquially, the Gödel sentence
is: ``This sentence cannot be proven in $T$’’. To avoid a logical contradiction, it would seem that one must interpret this sentence to actually be
true (an interpretation which was formally established later with Rosser's famous ``trick’’). So, by the  first incompleteness theorem, 
any logically consistent reasoning system is incomplete, not able to prove every true sentence without introducing contradictions.
 
Needless to say, this incompleteness theorem does not prevent mathematicians from having complete confidence that theorems they derive using the standard logical systems of mathematics are indeed true. The fact that if they wish to form a complete theory
there may be some other truth contradicting their theorems, a truth which their logical system cannot derive, 
does not prevent them from having this confidence. 
 
Going further though, Gödel’s \textit{second} incompleteness theorem says no sufficiently powerful logical system 
$T$ can contain a proof saying that $T$ is free of logical contradictions, no matter what precise mathematical axioms $T$ contains. 
One needs to introduce the lack of contradictions in $T$ as an additional axiom to establish that $T$ is free of contradictions. 
But that just means that we have replaced $T$ with a stronger logical system, $T’$, and $T’$ cannot prove that it is free of 
contradictions (even though it can prove that about $T$). We have an infinite regress, with no logical system $T$ able to validate all theorems
derived in $T$. But again, this does not prevent mathematicians from going about their business as usual, as though
there are no contradictions lurking out there that invalidate theorems they construct.
 
These two incompleteness theorems of Gödel are results in what is called `meta-mathematics', the mathematics of math itself. These two theorems (and many similar ones) tell us that mathematicians \textit{have no choice} but to  rely on what is called ``natural language’’ when they investigate meta-mathematics, a language that is not fully formal. Mathematicians \textit{have no choice} but to rely on ``pre-theoretical concepts of rationality"~\cite{dennett1981making}. But that is not sufficient reason for mathematicians to doubt theorems that
they collectively derive. Similarly, it is not sufficient reason to doubt the arguments in this essay --- yes,  
 
This phenomenon is rife in all of modern science, in fact, not just mathematics. For example, quantum mechanics tells us that there is 
a tiny-teeny probability that as you’re sitting on a sofa a lion will materialize out of thin air right next to you on the sofa. But this does not 
prevent physicists from confidently sitting in sofas without taking measures to protect themselves from possible attack by a neighboring feline
carnivore. The sole fact that some rather inconvenient fact \textit{may} be true (whether it be a lip-smacking lion suddenly staring you in the face 
or a mathematical truth that nullifies the beautiful theorem you think you just derived) does not cause modern scientists to pay any attention to it. 
The bar for paying attention to a fact is higher than it simply being possible. Results like those of Gödel are similar, just bereft of the tool of 
probability theory which we use to at least get a handle on these conundrums in quantum mechanics.
 
The argument of this essay, saying that arguments consisting only of finite sequences of symbols are almost unavoidably incomplete, 
has this very nature. It is very similar to a Gödel sentence, like the ones defined in his first incompleteness theorem.  
The implication of Gödel’s {second} incompleteness theorem is the same: yes, there may be aspects of reality 
that nullify this argument that finite 
sentences do not suffice. But that simple possibility of such an aspect of reality that would nullify this argument, by itself, 
provides no justification for disbelieving this argument. (Just like the possibility of contradictions in a proof presented
by a mathematician is not sufficient reason for other mathematicians to ignore that proof.) Perhaps
the most important lesson of the breakthroughs in our 20th century
SAM is that \textit{we have no choice} to live with this issue, if we \textit{ever} wish to grapple with 
the truly interesting issues in life, the deep issues, the issues lying beyond the nose
in front of our collective face.

This lesson immediately gives rise to the following question: 
\begin{enumerate}
\item 
[11.]
In standard formulations of mathematics, a mathematical proof 
is a finite sequence of ``well-formed sentences'', each of which is itself a finite
string of symbols. All of mathematics is a set of such proofs.
How would our perception of reality differ if, rather than just finite sequences of finite
symbol strings, the mathematics underlying our conception of reality 
was expanded to involve 
infinite sequences, i.e., proofs which do not reach their conclusion in finite time?
Phrased concretely, how would our cognitive abilities change if our brains could implement, or at least
encompass, super-Turing abilities, sometimes called ``hyper-computation'' (e.g., as proposed in computers that are on rockets moving arbitrarily
close to the speed of light~\cite{aaronson2005np})?

Going further,  as we currently conceive of mathematics, it is possible to embody all of its theorems, even those
with infinitely long proofs, in a single
countably infinite sequence: the successive digits of Chaitin’s omega~\cite{livi08}. (This is a consequence of the CT thesis.) How would mathematics differ
from our current conception of it if it were actually an uncountably
infinite collection of such countably infinite sequences rather than just one, a collection which could not be combined to
form a single, countably infinite sequence? Could we ever tell the difference? Could a being with super-Turing capabilities tell the difference, even if the CT
thesis is true, and even if we cannot tell the difference?

Going yet further, what would mathematics be if, rather than countable sequences
of finite symbol strings, it involved uncountable sequences of such symbol strings?
In other words, what if not all proofs were a discrete sequence of well-formed finite sentences, the successive sentences being indexed by counting integers, but rather some proofs were continuous sequences of sentences, the successive sentences being indexed by real 
numbers? Drilling further into the structure of proofs, what if some of the ``well-formed sentences" occurring in a proof’s sequence of 
sentences were not a finite sequence of symbols, but rather an uncountably infinite set of symbols?
If each sentence in a proof consisted of an
uncountably  infinite set of symbols, and in addition the sentences in the proof were indexed by a range of real numbers, then (formally speaking) the proof would be a curve  --- a one-dimensional object --- traversing a two-dimensional space. Going even further, what would it mean if somehow the proofs in God’s book~\cite{aigner2010proofs} 
were inherently \textit{multi}dimensional objects, not reducible to linearly ordered sequences of symbols, 
embedded in a space of more than two dimensions?

We can go even further than this set of issues. As mathematics is currently understood, the sequence of symbol strings in any proof 
must, with probability $1$, obey certain constraints. Proofs are the outcomes of
deductive reasoning, and so certain sequences of symbol strings are ``forbidden'', i.e., assigned probability $0$.
However, what if instead the sequences of mathematics were dynamically generated in a stochastic process, and therefore unavoidably random, with \textit{no} sequence assigned probability $0$~\cite{wolpert2024stochastic,wolpert2020noisy,christiano2013definability,dummett2000elements,franklin1987non}?
Might that, in fact, be how our mathematics has been generated? 
What would it be like to inhabit a physical universe whose laws could not be represented unless one used such a
mathematics~\cite{del2019physics,gisin2019indeterminism,gisin2020mathematical}?
Might that, in fact, be the universe that we do inhabit, but due to limitations in our minds, we cannot even conceive of all that extra 
stochastic structure, never mind recognize it? 

One has to be careful about the term ``conceive'' in these questions. There
\textit{has} been some research on infinitary logic, which
concerns logical systems in which the individual sentences are allowed to involve countably infinite sequences of symbols~\cite{stanford.encyclopedia.infinitary.logic}.  Similarly, cellular automata~\cite{wolfram1984cellular}
are models of computational systems that involve countably infinite sequences of symbols, where all of those
symbols matter in all iterations of the system (in contrast to the case of the tapes in Turing machines, which only
contain a finite number of non-blank symbols in each iteration). There have
also been recent work on using turbulent flow as (Turing-complete) computational systems~\cite{cardona2021constructing}.
Moreover, there is a long history of designing analog physical systems whose behavior exceeds what can
be calculated using Turing machines~\cite{pour1982noncomputability,siegelmann1998analog} or
more generally extending conventional models of computation to apply to real numbers~\cite{blum1998complexity}.
And of course, there \textit{has} been some preliminary investigation into the properties of
stochastic mathematical systems --- this work is cited above.

However, all of these examples 
are defined and reasoned about in journal papers, textbooks, computational theorem
provers and proof assistants, etc. In other words, they are themselves all formulated  
using finite sequences of finite sentences, each of which involves symbols from finite alphabets.
Even if these formalism concern SAM that does not have that finitary restriction,
when we define them, and when we reason about them,
we do so exclusively using SAM that does obey that finitary restriction.
After all, we \textit{have no choice} but to define and reason about these possible extensions to our SAM
using our own limited cognitive capabilities. Almost tautologically, we
cannot ``conceive'' of these possible extensions to SAM, in the sense of directly encompassing 
them in our minds. We can only construct structures that \textit{point} at those extensions; those
extensions themselves lie outside of us, outside the realm of what we can even conceive of.


So, as a final leap in this plexus of questions, we can ask: what would a 
mathematics be like whose very form 
could not be \textit{described} using a finite sequence of symbols from a finite alphabet, whose very definition
would involve infinitary structures of some sort, or perhaps even structures that are inherently stochastic?

\end{enumerate}

\noindent It's worth pointing out that
we can turn the (rather sharp) knife of question (11) for at least another half turn. Suppose that some future version of ourselves
{can}, somehow or other, construct and use one of the extensions of mathematics described in question (11),
an extension of mathematics that is one giant step beyond all we can currently even imagine.
Well, if that rather flabbergasting day ever were to dawn, as our future selves begin to cavort in its strange, new light, 
perhaps they will stumble upon misty vistas of \textit{further} leaps of kind of mathematics. 
Perhaps on that day, for the first time ever, the lineage of \textit{Homo Sapiens} will 
wonder about kinds of reasoning that cannot be adumbrated by our current selves,
that can only be even dimly perceived under {the} strange glow cast by one of the extensions of mathematics described in question (11).
Perhaps our children will stumble upon a sort of question (11) built on top of question (11). 
An extension of mathematics that is \textit{two} giant steps beyond the kinds of mathematics we can currently even
formulate. 

And no, I have no idea of what that last sentence means, not really. That's the point in fact.

Okay, bringing the discussion back down to earth (or at least down to the solar system),
is the farrago of issues raised by question (11), those tight twists in the collective underwear of our jejune minds,
are they all much ado about nothing? Is there any reason to suppose that mathematical structures
we cannot conceive of might be important features of (for lack of a better word) ``reality''?

In a certain sense, \textit{by definition}, we can never definitively answer `no' to this question. (We can never rule
out the possibility that something beyond what our minds can conceive somehow affects that which we perceive.)
On the other hand, due to the limitations
on us that motivated question (11) in the first place, \textit{in practice} we can never answer `yes'. So
it seems that we cannot definitively answer question (11). Or at least, that is the conclusion concerning our current reasoning
abilities that our current reasoning leads us to.

Phrasing things that way leads us
to an issue that was briefly discussed above, just after question (9): How might the set of what-we-can-conceive 
evolve in the future? Suppose that, indeed, the kinds of knowledge concerning mathematical and physical reality that our current brains 
can conceive of having is a proper subset of all of the kinds of knowledge there are. In other words, suppose the set of 
what-can-be-known-but-not-even-conceived-by-us is non-empty. Is there any way we can know that?

\begin{enumerate}
\item [12.]	Is there any way that we imagine testing --- or at least gaining insight into --- whether our SAM can, in the future, capture all of physical reality?
If not, is there any way of gaining insight into how much of reality is forever beyond our ability to even conceive of? In short, what can we ever
know about the nature of that which we cannot conceive of?
\end{enumerate}

Waxing meta for a moment, note that
question (12) indirectly highlights a tension in the arguments I've been using throughout this essay.
As was also pointed out just after question (9), the tools that a paramecium can conceive
of using to come to grips with the full nature of physical reality are extraordinarily limited (perhaps involving chemotaxis
and the like). 
I gave as an example the fact that a paramecium cannot
even conceive of \textit{questions and their answers}. Yet we, their intellectual successors,
feel that questions and answers are an indispensable tool for grappling with reality.
It's hard not to see the immediate analogy. A paramecium is wrong in its (imputed, implicit) view
of what is important to comprehend physical reality; as is obvious to any creature more
intelligent than a paramecium, sensing directions of chemical flows is pretty much irrelevant. And in exact analogy,
\textit{we} may be wrong in \textit{our} view of what is important to comprehend physical reality --- as would be obvious
to a creature more intelligent than \textit{we} are. So in particular, asking questions --- precisely as I am doing in this essay ---
might be completely wrong-headed, irrelevant to engaging cognitively with physical reality.

Yep, that's right. That concern is almost tautological in fact.
Nonetheless, asking questions has gotten us this far --- and I for one cannot think (!) of
any other way to make progress. So, let's run with them a bit longer: What can we say 
in direct answer to question (12)?

First, from a certain perspective, question (12) might appear to
be a scientific version of a conspiracy theory, writ large. One
might argue that it is no different in kind from the fact that: you can never prove that ghosts don’t exist, either theoretically or empirically; that Marduk doesn't really pull the
strings in human affairs, nor does an Abrahamic deity; that it is not actually the believers in QAnon who are members of a secret cabal of Satan-worshiping cannibalistic
pedophiles running a global child sex-trafficking ring and engaged in a massive (highly successful) disinformation campaign.

Adopting this perspective, one might say, ``Sure, yes, it may be that there are unconceivable aspects to reality. We can’t somehow 
prove that there 
aren't. But by definition, we could never know such unconceivables --- could never investigate them empirically --- so who cares?'' Isn't the ending 
question of this essay about can-never-be-conceived aspects of reality just a pompously inflated version of the platitude that you can’t prove a 
negative? The famous scalpel of Occam suggests you should expunge consideration of such a proposition from your intellectual to-do list. Just as the 
scalpel denigrates the proposition that there is an Abrahamic deity in the sense that is defended by Deists, deriding that proposition as vacuous, said 
scalpel also suggests that the final question of this essay is vacuous, with no consequences (by definition), and therefore unworthy of contemplation.

However, even if there are aspects of reality that will forever lie beyond our ken, that does not
mean that might have consequences which have major empirical consequences for us. For
example, consider those aspects of our physical universe that will forever characterize as ``random''
(in the case of quantum mechanics) or ``beyond our ability to calculate'' (e.g., due to intrinsic
chaos, or simply because the calculation would require more iterations of a Turing machine than
could be run before the heat death of the universe), or ``arbitrary'' (e.g., the physical constants of nature,
or the details of the universe's boundary conditions). There may even be physical phenomena that
are neither deterministic nor stochastic, e.g., computationally undecidable, but whose existence has major
empirical consequences for us. 

On the other hand, it may also be that if we simply had
sufficient cognitive capability, the mystery of all such phenomena would be resolved for us, and
we would be able to understand / predict them in full. So even if we cannot answer the question I am posing
in this essay, that answer might have hugely consequential implications for what we observe around us.

Moreover, there are at least five reasons to suspect that we actually \textit{can} 
find the answer to (some aspects of) the question I am posing in this essay, which salvages the 
question from the realm of such vacuities:
\begin{enumerate}
\item[i)] Most prosaically, we may be able to make some inroads into what is currently beyond our ability to conceive of we can ever somehow construct a super-Turing computer~\cite{aaronson2013philosophers} 
and exploit it to consider the question of what knowledge can never be conceived by us, thereby breaking free of the strictures of the CT thesis. 

\item[ii)] More speculatively, our cognitive abilities will presumably continue to grow, e.g., by our constructing new 
kinds of external information retrieval and processing systems (earlier examples of which are books, then 
computers, and now the web), or by our building neurobiological prostheses, directly connected to our
brains. These new abilities might at least be able to establish the existence of what we can never conceive, through some 
observational / simulational / theoretical / {who knows?} process. In other words, it may be that the feedback loop between our extended minds and our technology does let us break free of the evolutionary accident --- of the fact that our minds were formed for the Veldt --- 
at least in certain regards. It may be that the answer to question 4 above is ``no" ---  and this may have consequences.

\item[iii)] Suppose we figure out how to communicate with ``alien intelligences'' here on Earth. For example, we may figure
out how to communicate with Cetaceans, or with the (still very speculative) ``wood-wide web'' of communities of
trees in forests who communicate with another both directly and via mycorrhizal networks. One could imagine
that these terrestrial alien intelligences might be able to answer the question I am posing.

\item[iv)] Suppose we encounter extraterrestrial intelligence, e.g., by plugging into some vast galaxy-wide web of interspecies discourse, containing in particular some cosmic version of Stack Exchange. To answer the question --- to determine whether there are aspects of physical reality that are knowable but that we humans cannot even
conceive of that kind of knowledge --- might require nothing more than our posing that question to the cosmic Stack Exchange, and then reading through the answers that get posted.

\item[v)] Consider our evolutionary progeny --- not just future variants of our species that evolve from us via conventional Neo-Darwinian
 (epi)genetic mutation, crossover, and selection, but future members of any organic species 
that we consciously design.\footnote{To give a simple example, it seems highly likely that within
the next few decades very many people will start to apply gene-editing techniques (e.g., via CRISPR) to the human germ
line, in order to make babies with super-human intelligence. }

 It seems quite
 likely that the minds of such successors of ours will have a larger set of things that can be conceived than we do. After all, the decades-long
 experiments conducted in Richard Lenski’s lab~\cite{blount2018contingency} 
show that even without any consciously directed intervention (relying solely on the 
proverbial Neo-Darwinian ``blind watchmaker" to guide the evolution,
albeit a sped-up watchmaker), there is unceasing growth in adaptive fitness. There is no reason to believe that this would not also apply to
cognitive ``fitness." (See the discussion above of a time-series analysis of the set of cognitive capabilities of terrestrial species.)

In addition, the successors that we design would be an evolutionary first: a species directly constructed in a goal-oriented, Lamarkian manner. This novelty in how our descendant species will be constructed may mean that their cognitive ``genotype" will have jumped out of our current metastable
state~\cite{aguirre2018networked,manrubia2021genotypes}. 
The jump from us to them may be a massive punctuated equilibrium, overcoming the limitations imposed on us by the frozen accidents in our evolutionary past.

\item[vi)] This all holds just as well for any progeny we design that are inorganic, i.e., for future Artificial Intelligences. Indeed,  consider
``smart contracts'', which are programs that are implemented via block chains that
provide a means for multiple AIs to enter contracts with one
another. A crucial property of smart contracts is that are both complete and fully binding 
(in the terminology of economics), even without the presence of a judicial system 
enforcing those contracts. Moreover, currently at least, smart contracts are publicly visible by default. 

Combining these traits means that in general it is possible for  two AIs, $A$ and $B$, to
enter a smart contract and for a third AI, $C$, to observe that contract before it is implemented, and 
extort money from one of those two AIs (suppose it's $A$).
The way they do this is completely legal: they make a threat to $A$ of the following form: ``Give me $\$ x$ now,
before this contract is executed, or else \textit{I will give you
a certain amount of money after you and $B$ have executed your contract}''.\footnote{The key for such a threat
to work is that the amount of money that $C$ threatens to give to $A$ varies depending on what $A$ chooses to do when the
contract is implemented --- and $B$ knows this. See~\cite{bono_wolpert-game-mining,bono_wolpert_game_mining_2014,ramirez2023game}.}

Going beyond this scenario involving three AIs,
what can we predict about how whether there are other such bizarre situations that can arise, for other smart contracts?
Well, smart contracts are {Turing complete}. (Loosely speaking, they are arbitrary Python code.) This means that
according to the Church-Turing thesis, it is \textit{uncomputable by we humans} what might happen when AIs start to
engage in such smart contracts with one another~\cite{biderman2020magic,dargaj2020complete,churchill2019magic,livi08}. In other words, quite soon now, the networks of AIs interacting with
one another via smart contracts \textit{will be outside of the set of mathematical problems that humans can solve,
even in theory}. 

This prospect should not be very surprising. 
Recall that in 2010, when a relatively small number of very dumb
trading bots interacted with one another via a highly constrained set of protocols, the result was the flash crash, wiping 1000 points off
the Dow Jones stock index in a matter of minutes --- and then recovering just as fast. Note that \textit{we still have no formal understanding
of just what happened in the flash crash~\cite{johnson2013abrupt}.} 
(That's why with the exception of some crude restrictions that were imposed in Singapore
in response to a later crash, no regulations have been implemented to prevent bot-caused flash crashes from
occurring again.) What will future versions of ``flash crashes'' be, when we replace those simple bots with future AIs, and allow
them to interact arbitrarily, via smart contracts? These future AI-ecosystem eruptions will very likely be far, far beyond our understanding ...
and in many ways may constitute intelligences, far ``advanced'' above our own.


So in a very real sense, the old idea that there is a technological ``singularity''
in our near future~\cite{vinge1993coming}, at which time our machines will start to behave in
ways that are beyond our ability to predict even in theory, is completely correct. But the singularity will not arise  
by having monolithic
systems, like AGI, or generative AIs, get more and more powerful. Such thinking, focused on
single systems operating in isolation, is so retro, so \textit{twentieth century}, for lack of a 
better expression. If there's one lesson of the last
several decades of advances in computer sciences, it's that leaps arise in \textit{distributed} systems, not
monolithic ones. Yes, the singularity will occur --- but it will be distributed.
%
%
\end{enumerate}

The key issue is not whether our evolutionary children will be organic (involving modification to biological germ lines)
or inorganic (involving the interaction of multiple AIs), or even a mixture of the two. It's the fact that those little kiddies 
of ours will likely have cognitive abilities far advanced of ours. And they will likely
arrive, with those attendant cognitive abilities, within 50 --- 100 years. 

Presumably we will go extinct soon after their arrival (like all good parents making way for their children). So, on our way out the door, as we gaze up at our successors in open-mouthed wonder, as one of our last acts:

\begin{quotation}
We can simply ask our question of them.
\end{quotation}

\section*{Acknowledgments:} I would like to thank David Kinney, Mikhail Prokopenko, as well as Daniel Dennett 
for interesting conversation on these issues. This work was supported by funding
from the Santa Fe Institute.

\bibliographystyle{amsplain}
\bibliography{/Users/davidwolpert/Dropbox/BIB/refs,/Users/davidwolpert/Dropbox/BIB/refs.main.1.BIB.DIR}

\end{document}